\documentstyle[12pt]{article}
\title{\bf
Discrete symmetry breaking and restoration at finite temperature
in 3D Gross-Neveu model}
\author{
{\bf 
Bang-Rong Zhou}\thanks{Regular Associate of The Abdus Salam ICTP.}  \\
\normalsize Department of Physics, Graduate School at Beijing \\
\normalsize University of Science and Technology of China,
  Academia Sinica, Beijing 100039, {\bf China} \\
\normalsize and  \\
\normalsize The Abdus Salam International Centre for Theoretical Physics,
Trieste 34100, {\bf Italy} \\
}
\date {}
\begin{document}
\hoffset = -1 truecm
\voffset = -2 truecm
\baselineskip = 12pt
\maketitle

\begin{abstract}
Dynamical spontaneous breaking of some discrete symmetries including special
parities and time reversal and their restoration at finite temperature $T$ are
researched in 3D Gross-Neveu model by means of Schwinger-Dyson equation in the
real-time thermal field theory in the fermion bubble diagram approximation.
When the momentum cut-off $\Lambda$ is large enough, the equation of critical
chemical potential $\mu_c$ and critical temperature $T_c$ will be
$\Lambda$-independent and identical to the one obtained by auxialiary scalar
field approach.  The dynamical fermion mass $m\equiv m(T,\mu)$, as the order
parameter of symmetry breaking, has the same $(T_c-T)^{1/2}$ behavior at
$T\stackrel{<}{\sim} T_c$ as one in 4D NJL-model and this shows the second-order phase
transition feature of the symmetry restoration at $T>T_c$. It is also proven
that no scalar bound state could exist in this model.
\end {abstract}
\newpage

\indent The Gross-Neveu (GN) model [1] in 2+1 space-time dimension is a model
with four-fermion interactions. As a well-known and applicable model to codensed
matter physics, it has been extensively researched [2-6]. The coventional
approach is to introduce an auxialiary scalar field, then to replace the
four-fermion interactions by the Yukawa couplings between the auxialiary scalar
field and the fermion fields. Alternatively, one can also use Schwinger-Dyson
equation directly to research the fermion condensates induced by the
four-fermion interactions and relevant dynamical behavier. In the latter
approach, no effective scalar field is introduced and, as far as some problems,
for example, the calculation of the propagators for the bound states composed of
fermions at finite temperature,  are concerned,  this approach could be more
straightforward and convenient.
On the other hand,  almost all the discussions about the finite
temperature behavior of the model were conducted in the imaginary-time formalism
of thermal field theory. However, we also want to examine the application of the
real-time formalism of thermal field theory to the model.  Therefore, in this
letter, instead of using the auxialiary scalar field and the imaginary-time
thermal field theory, we will research the 3D GN model at finite
temperature by means of Schwinger-Dyson equation in the real-time thermal field
theory.  All the discussions will be made in the fermion bubble diagram
approximation.  This approximation amounts to the leading order of $1/N$
expansion and is sometimes also called the mean field approximation. Since
3D GN model is renormalizable in $1/N$ expansion [5] and in 2+1 dimension,
there is no the kink effect which could appear in a 1+1 dimension system
[7], the mean field approximation can be considered as a reliable one.  The
results will also be compared with the previous works' ones based on the
auxialiary field and the imaginary-time thermal field theory. \\
\indent The Lagrangian of the model can be written by
\begin{equation}
{\cal L}(x)=\sum_{k=1}^N\bar{\psi}^k(x)i\gamma^{\mu}\partial_{\mu}\psi_k(x)+
              \frac{g}{2N}\sum_{k=1}^N[\bar{\psi}^k(x)\psi_k(x)]^2.
\end{equation}
\noindent In 3 dimensions, the coupling constant $g$ has mass dimension -1 thus
the model is perturbatively non-renormalizable. However, this fact does not
affect our discussions based on the $1/N$ expansion. The
$\gamma^{\mu}\ (\mu =0,1,2)$ can be taken as $2\times 2$ matrices and related
to the Pauli matries $\sigma^i (i=1,2,3)$ by
\begin{equation}
\gamma^0=\sigma^3=\left(\matrix{
                         1 &0 \cr
                         0 &-1 \cr} \right), \ \
  \gamma^1=i\sigma^1=\left(\matrix{
                         0 &i \cr
                         i &0 \cr} \right), \ \
  \gamma^2=i\sigma^2=\left(\matrix{
                         0 &1 \cr
                         -1&0 \cr} \right),
\end{equation}
\noindent which satisfy the Dirac algebra
\begin{equation}
\left\{\gamma^{\mu}, \gamma^{\nu}\right\}=2g^{\mu \nu}, \ \
  \mu, \nu =0,1,2.
\end{equation}
\noindent In this representation $\psi_k(x)$ is two-component complex spinor
with $N$ "color" degrees of freedom. Notice that there is no $2\times 2$ matrix
which anti-commutes with all the three $\gamma^{\mu}$, i.e. no "$\gamma_5$"
matrix exists in 3 dimensions. Consequently ${\cal L}(x)$ has no chiral
symmetry, though $\psi_k(x)$ are zero-mass fields. However, it is easy to verify
that the action of system is invariant for the discrete transformation (omitting
the index $k$ of $\psi$)
\begin{equation}
\psi^1(x)\stackrel{Z_2}{\longleftrightarrow}\psi^2(x), \ \ {\rm if} \
\psi(x)=\left(\matrix{
              \psi^1(x) \cr
              \psi^2(x) \cr}\right),
\end{equation}
\noindent the special ${\cal P}_1$ and ${\cal P}_2$ parity transformations
\begin{equation}
\psi(t, x^1, x^2)\stackrel{{\cal P}_1}{\longrightarrow}
\gamma^1\psi(t,-x^1, x^2),
\end{equation}
\begin{equation}
\psi(t, x^1, x^2)\stackrel{{\cal P}_2}{\longrightarrow}
\gamma^2\psi(t,x^1, -x^2),
\end{equation}
\noindent and the time reversal
\begin{equation}
\psi(t, x^1, x^2)\stackrel{{\cal T}}{\longrightarrow}
\gamma^2\psi(-t,x^1, x^2).
\end{equation}
\noindent We indicate that the mass term
$-m\bar{\psi}(x)\psi(x)$ (if it exists) is not invariant under $Z_2$,
${\cal P}_1$, ${\cal P}_2$ and ${\cal T}$. Hence, if the four-fermion interactions could lead to generation
of a non-zero dynamical mass of the $\psi$-fields, then the above $Z_2$,
${\cal P}_1$, ${\cal P}_2$ and ${\cal T}$ discrete symmetries will be
spontaneously broken. Since the broken symmetries are discrete, no Goldstone
bosons are expected to appear.  In the following we will research
generation of the dynamical mass of $\psi(x)$ fields and its finite temperature
behavior in the fermionic bubble graph approximation.  This approximation, as
the leading order of the $1/N$ expansion, will be able to give us a clear
insight into the breaking and restoration of above descrete symmetries.
 \\
\indent At zero-temperature $T=0$, assume the four-fermion scalar interactions
$g\sum_{k=1}^N{(\bar{\psi}^k\psi_k)}^2/2N$ can lead to the fermion condensates
$\sum_{k=1}^N\langle\bar{\psi}^k\psi_k\rangle\neq 0$, then we will get the
dynamical fermion mass $m(0)=-(g/N)\sum_{k=1}^N\langle\bar{\psi}^k\psi_k\rangle$
which satisfies the gap equation
\begin{equation}
1=2g\int\frac{id^3l}{{(2\pi)}^3}\frac{1}{l^2-m^2(0)+i\varepsilon}
\end{equation}
\noindent After Wick rotation, angular integration and introduction of
3D Euclidean momentum cut-off $\Lambda$, Eq. (8) becomes
\begin{equation}
1=\frac{g\Lambda}{\pi^2}\left[1-
       \frac{m(0)}{\Lambda}\arctan\frac{\Lambda}{m(0)}\right].
\end{equation}
\noindent We indicate that since $\Lambda$ corredponds to the square root of the
3d squared Euclidean momentum, its introduction does not explicitly break the
discussed discrete symmetries including ${\cal P}_1$, ${\cal P}_2$ and
${\cal T}$. It is seen from Eq.(9) that the necessary condition for formation
of the fermion condensates is sufficiently strong four-fermion coupling
$g$ that $g\Lambda/\pi^2>1$, similar to the one in 4D NJL-model case [8]. In
both cases, one can consider the four-fermion couplings as an effective
interactions below some high momentum scale $\Lambda$.  For the sake of
satisfying the gap equation  Eq. (8), the fine-tuning of the coupling constant
$g$ is also required.\\
\indent When $T\neq 0$, we must replace the vacuum expectation value
$\sum_{k=1}^N\langle\bar{\psi}^k\psi_k\rangle$ by the thermal expectation
value $\sum_{k=1}^N{\langle\bar{\psi}^k\psi_k\rangle}_T$. In the real-time
thermal field theory, this implies the substitution of the fermion propagator
[9,10]
\begin{equation}
\frac{i}{\not\!{l}-m+i\varepsilon}\rightarrow
  \frac{i}{\not\!{l}-m+i\varepsilon}-2\pi\delta(l^2-m^2)
  (\not\!{l}+m)\sin^2\theta(l^0,\mu)
\end{equation}
\noindent with
\begin{equation}
\sin^2\theta(l^0, \mu)=\frac{\theta(l^0)}{\exp[\beta(l^0-\mu)]+1}
                        +\frac{\theta(-l^0)}{\exp[\beta(-l^0+\mu)]+1},
\end{equation}
\noindent where $m\equiv m(T, \mu)$ is the dynamical fermion mass at finite
temperature $T$ and finite chemical potential $\mu$ and the denotation
$\beta=1/T$. As a result, the gap equation at $T\neq 0$ becomes
\begin{equation}
1=gI
\end{equation}
\begin{eqnarray}
I&=&2\int\frac{d^3l}{{(2\pi)}^3}\left[
   \frac{i}{l^2-m^2+i\varepsilon}-2\pi\delta(l^2-m^2)\sin^2\theta(l^0,\mu)
   \right]  \\ 
 &=&\frac{\Lambda}{\pi^2}\left\{
   1-\frac{m}{\Lambda}\arctan\frac{\Lambda}{m}-\frac{\pi T}{2\Lambda}
   [I_2(y,-r)+I_2(y,r)]\right\}
\end{eqnarray}
\noindent where we have used the denotations
\begin{equation}
I_2(y,\mp r)=\int_{0}^{\infty}\frac{dx x}{\sqrt{x^2+y^2}}
\frac{1}{\exp(\sqrt{x^2+y^2}\mp r)+1}=\ln\left[1+e^{-(y\mp r)}\right]
\end{equation}
\noindent with
\begin{equation}
x=\beta |\stackrel{\rightharpoonup}{l}|, \ y=\beta m \ {\rm and} \
  r=\beta \mu.
\end{equation}
\noindent Considering the gap equation (9) at $T=0$, Eq.(12) may be changed
into
\begin{equation}
m(0)\arctan\frac{\Lambda}{m(0)}= m \ \arctan\frac{\Lambda}{m}+\frac{\pi}{2}f(T)
\end{equation}
\noindent with
\begin{equation}
f(T)=T\ln\left\{\left[1+e^{-(m-\mu)/T}\right]
\left[1+e^{-(m+\mu)/T}\right]\right\}.
\end{equation}
\indent It is easy to see that no matter $m-\mu\geq 0$ or $m-\mu<0$ we always
have $f'(T)=df(T)/dT>0$, i.e. $f(T)$ is a monotone increasing function of
temperature $T$. To keep the left-handed side of Eq.(17) to be a constant,
the first term in the right-handed side must decreases as T increases and
finally has $m\rightarrow 0$ in it at a critical temperature $T_c$ and a
critical chemical potential $\mu_c$.  Hence taking $m=0$ in Eq.(17) we will
obtain the equation to determine $T_c$ and $\mu_c$
\begin{equation}
m(0)\arctan\frac{\Lambda}{m(0)}=\frac{\pi}{2}T_c\left[\ln (1+e^{\mu_c/T_c})+
                        \ln(1+e^{-\mu_c/T_c})\right].
\end{equation}
\noindent Here the explicit momentum cut-off $\Lambda$ is not important.  In
fact, as long as the ratio $\Lambda/m(0)\geq 10^4$, then the value of
$\arctan[\Lambda/m(0)]$ is almost equal to $\pi/2$. Assume that is the case
then the momentum cut-off $\Lambda$ will disappear from the equation and we
obtain
\begin{equation}
m(0)=T_c\left[\ln (1+e^{\mu_c/T_c})+
                        \ln(1+e^{-\mu_c/T_c})\right].
\end{equation}
\noindent We indicate that the $\mu-T$ criticality equation (20) comes from the
two-point function of frmions which corresponds to $O(1)$ order in $1/N$
expansion, hence no explicit $N$-dependence is contained in it. It follows from
Eq.(20) that
\begin{equation}
\matrix{T_c/m(0)=1/2\ln 2, & {\rm if} \ &\mu_c=0  \cr
          \mu_c/m(0)=1,  & {\rm if} \ &T_c=0 \cr}
\end{equation}
\noindent If scaled in $m(0)$ by defining $u_c=\mu_c/m(0)$ and $t_c=T_c/m(0)$
then the equation (20) of criticality curve of chemical potential and
temperature can be rewritten as
\begin{equation}
u_c=1+2t_c\ln[(1+\sqrt{1-4e^{-1/t_c}})/2].
\end{equation}
\noindent These results obtained under the assumption $\Lambda/m(0)\gg 1$ are
identical to the ones given by effective scalar field approach in the leading
order of $1/N$ expansion [5] and also cosistent with the calculations of Ref.
[4]. \\
\indent By means of the dynamical fermion mass, we may have more clear
understanding of the feature of symmetry restoration phase transition. Assuming
$\Lambda/m(0)\gg 1$ and in view of $m\leq m(0)$, we will have
$\arctan[\Lambda/m(0)]\simeq \arctan[\Lambda/m]\simeq \pi/2$.  Thus Eq.(17)
will become $\Lambda$-indepedent, i.e.
\begin{equation}
m(0)= m+T\left\{\ln \left[1+e^{-(m-\mu)/T}\right]+
                  \ln \left[1+e^{-(m+\mu)/T}\right]\right\}.
\end{equation}
\noindent When $T\stackrel{<}{\sim} T_c$, we have $m/T\ll 1$ thus Eq.(23) can be
transformed into
\begin{eqnarray}
m(0)&=&T\left[\ln\left(1+e^{\mu/T}\right)+\ln\left(1+e^{-\mu/T}\right)\right]+
      T\ln \left[1+\left(e^{m/T}-1\right)+
                 \frac{{\left(e^{m/T}-1\right)}^2}{\left(1+e^{\mu/T}\right)
                 \left(1+e^{-\mu/T}\right)}
           \right]-m  \nonumber \\
&\simeq &T\left\{
      \left[\ln\left(1+e^{\mu/T}\right)+\ln\left(1+e^{-\mu/T}\right)\right]+
      \frac{m^2}{2T^2[1+\cosh(\mu/T)]}+{\cal O}\left(\frac{m^3}{T^3}\right)
          \right\},
\end{eqnarray}
\noindent where we have made the power series expansion of $[\exp(m/T)-1]$ in $m/T$
and kept only the terms to the order ${\cal O}(m^2/T^2)$.  Combining Eq.(24)
with Eq.(20) (in which $\mu_c$ is replaced by $\mu$) we obtain
\begin{equation}
m^2=2T\left(1+\cosh \frac{\mu}{T}\right)\left\{
      T_c\ln \left[2\left(
      1+\cosh \frac{\mu}{T_c}\right)\right]-
      T\ln \left[2\left(1+\cosh \frac{\mu}{T}\right)\right]
                                          \right\}.
\end{equation}
\noindent By means of the approximate expression at $T\stackrel{<}{\sim} T_c$
\begin{eqnarray}
\ln \left[2\left(1+\cosh \frac{\mu}{T}\right)\right]&=&
\ln \left[2\left(1+\cosh \left\{\frac{\mu}{T_c}+\frac{\mu(T_c-T)}{T_cT}
                         \right\} \right)\right] \nonumber \\
&\simeq &\ln\left[2\left(1+\cosh \frac{\mu}{T_c}\right)\right]+
          \frac{\mu(T_c-T)}{T_cT}\tanh\frac{\mu}{2T_c}
\end{eqnarray}
\noindent Eq.(25) may be changed into
\begin{equation}
m^2=2T\left(1+\cosh \frac{\mu}{T}\right)\left\{
      \ln \left[2\left(1+\cosh \frac{\mu}{T_c}\right)\right]-
      \frac{\mu}{T_c}\tanh \frac{\mu}{2T_c}\right\}(T_c-T) \ \
      {\rm when} \ T\stackrel{<}{\sim} T_c
\end{equation}
\noindent We see that the dynamical fermion mass $m$, as the only order
parameter of the discrete symmetry breaking, will have $(T_c-T)^{1/2}$
behavior at $T\stackrel{<}{\sim} T_c$. This situation is similar to
the one in 4 dimension NJL-model [10].  Hence, based on the same arguments given
in Ref. [10], we can
conclude that in the 3D Gross-Neveu model the discrete symmetry restoration at
 $T>T_c$ will also be a secon-order phase transition. \\
 \indent In this model, no continuous symmetry breaking means no Goldstone boson
appearing.  An interesting question is that whether any (massive or massless)
scalar bound state could exist. Such scalar bound state can be thought as the
equivalent of a "Higgs" particle appearing in spontaneous breaking of a
continuous symmetry.  In present case, it could be the configuration
$\sum_{k=1}^{N}({\bar{\psi}}^k\psi_k)$. However, we will prove that the answer
to the above question is negative in 3D Gross-Neveu model.
Based on the method taken in Ref.[11] in the real-time formalism of thermal
field theory, we can obtain from the scalar four-point function of fermions
the formal propagator for the "scalar bound state" expressed by
\begin{equation}
\Gamma_S(p)=-i/N(p^2-4m^2+i\varepsilon)\left[
              K(p)+H(p)-iS(p)-\frac{R^2(p)}{K(p)+H(p)+iS(p)}\right],
\end{equation}
\noindent where the functions
\begin{eqnarray*}
K(p)&=&\frac{1}{4\pi^2}\int_{0}^{1}\frac{dx}{[m^2-p^2x(1-x)]^{1/2}}
       \arctan\frac{\Lambda}{[m^2-p^2x(1-x)]^{1/2}}  \\
    &\simeq &\frac{1}{8\pi \sqrt{p^2}}\ln
             \frac{4m^2+4m\sqrt{p^2}+p^2}{4m^2-p^2}, \ {\rm if} \
             \arctan\frac{\Lambda}{[m^2-p^2x(1-x)]^{1/2}}\approx \frac{\pi}{2} \
             {\rm and}  \ 0<p^2<4m^2,\\
\end{eqnarray*}
$$
H(p)=2\pi \int \frac{d^3l}{(2\pi)^3}\left\{
       \frac{(l+p)^2-m^2}{[(l+p)^2-m^2]^2+\varepsilon^2}+(p \to -p)\right\}
       \delta(l^2-m^2)\sin^2\theta(l^0,\mu),
$$
$$
S(p)=2\pi^2 \int \frac{d^3l}{(2\pi)^3}\delta(l^2-m^2)\delta[(l+p)^2-m^2]
       \left[\sin^2\theta(l^0,\mu)\cos^2\theta(l^0+p^0,\mu)\right. $$ $$
       \left.+ \cos^2\theta(l^0,\mu)\sin^2\theta(l^0+p^0,\mu)\right],
$$
\begin{equation}
R(p)=\pi^2 \int \frac{d^3l}{(2\pi)^3}\delta(l^2-m^2)\delta[(l+p)^2-m^2]
       \sin2\theta(l^0,\mu)\sin2\theta(l^0+p^0,\mu).
\end{equation}
\noindent It is indicated that when deriving Eq. (28) we have used Eq. (12).
As a result, the $\Lambda$-dependent divergent sectors in $\Gamma_S(p)$ have
cancelled each other and this is just a peculiarity in the leading order of
$1/N$ expansion.
Formally $p^2=4m^2$ seems to be a mere simple pole of $\Gamma_S(p)$,
but this is not true. First of all we notice that when $p^2\to 4m^2$,
\begin{equation}
\lim \limits_{p^2\to 4m^2}K(p)\rightarrow \infty
\end{equation}
\noindent and $K(p)$ has a logarithmic singularity.  Next, we can write
$$
H(p)=\frac{1}{8\pi^2}\int \frac{d^2l}{\omega_l}\{
   (\frac{p^2+2\omega_lp^0-2|\stackrel{\rightharpoonup}{l}|
   |\stackrel{\rightharpoonup}{p}|\cos\varphi}
   {[p^2+2\omega_lp^0-2|\stackrel{\rightharpoonup}{l}|
   |\stackrel{\rightharpoonup}{p}|\cos\varphi]^2+\varepsilon^2} +
         \frac{p^2-2\omega_lp^0+2|\stackrel{\rightharpoonup}{l}|
   |\stackrel{\rightharpoonup}{p}|\cos\varphi}
   {[p^2-2\omega_lp^0+2|\stackrel{\rightharpoonup}{l}|
   |\stackrel{\rightharpoonup}{p}|\cos\varphi]^2+\varepsilon^2})
$$
\begin{equation}
   \cdot \frac{1}{\exp[\beta(\omega_l-\mu)]+1}+
   (p^0\to -p^0, \mu\to -\mu)\}, \ \omega_l={(
   {\stackrel{\rightharpoonup}{l}}^2+m^2)}^{1/2}.
\end{equation}
\noindent The angular integrations in Eq. (31) are of the following general
forms and results:
\begin{equation}
\int_0^{2\pi}d\varphi \frac{1}{b+c\cos\varphi\pm\varepsilon^2} =0,
\end{equation}
\noindent where $b$ and $c$ are constants.  Hence the function $H(p)=0$.  As
for $S(p)$ and $R(p)$, when $p^2=4m^2$ each of them may be generally expressed
by
\begin{eqnarray*}
A(p)|_{p^2=4m^2}&=&\int d^3l \ \delta(l^2-m^2)\delta[(l+p)^2-m^2]
                  f(l^0,p^0,\mu)|_{p^2=4m^2}   \\
                &=&\int_0^{\infty}\frac{d|\stackrel{\rightharpoonup}{l}|
                   |\stackrel{\rightharpoonup}{l}|}{4\omega_l}
                   \int_0^{2\pi} d\varphi  \ \delta(2m^2-\omega_lp^0-
                   |\stackrel{\rightharpoonup}{l}||\stackrel{\rightharpoonup}{p}|
                   \cos\varphi)f(-\omega_l,p^0,\mu),
\end{eqnarray*}
\begin{equation} \end{equation}
\noindent where $f(l^0,p^0,\mu)$ is a finite function. In Eq.(33), the condition
$|\cos\varphi| \leq 1$ limits
$|\stackrel{\rightharpoonup}{l}|$ to the only possible value
$|\stackrel{\rightharpoonup}{l}|=|\stackrel{\rightharpoonup}{p}|/2$. Let
$\tan(\varphi/2)=t$, then the integration of the angle $\varphi$ in Eq. (33)
can be written by
\begin{eqnarray}
B&=&\left(\int_0^{\pi}+\int_{\pi}^{2\pi}\right)d\varphi \
  \delta(2m^2-\omega_lp^0-|\stackrel{\rightharpoonup}{l}|
                         |\stackrel{\rightharpoonup}{p}|\cos\varphi)\nonumber \\
&=&2\int_0^{\infty}dt\left\{\delta[(2m^2-\omega_lp^0+
    |\stackrel{\rightharpoonup}{l}||\stackrel{\rightharpoonup}{p}|)t^2+
(2m^2-\omega_lp^0-|\stackrel{\rightharpoonup}{l}||\stackrel{\rightharpoonup}{p}|)]
\right. \nonumber \\
&&\left.+\delta[(2m^2-\omega_lp^0-
    |\stackrel{\rightharpoonup}{l}||\stackrel{\rightharpoonup}{p}|)t^2+
(2m^2-\omega_lp^0+|\stackrel{\rightharpoonup}{l}||\stackrel{\rightharpoonup}{p}|)]
\right\}.
\end{eqnarray}
\noindent Since the $\delta$-functions have non-zero values only if
$|\stackrel{\rightharpoonup}{l}|=|\stackrel{\rightharpoonup}{p}|/2$, we can set
\begin{equation}
|\stackrel{\rightharpoonup}{l}|=
(\frac{1}{2}\pm \varepsilon)|\stackrel{\rightharpoonup}{p}|, \ \
\varepsilon =0_+
\end{equation}
\noindent and obtain
$$
2m^2-\omega_lp^0+|\stackrel{\rightharpoonup}{l}||\stackrel{\rightharpoonup}{p}|
  \simeq \frac{\varepsilon^2{\stackrel{\rightharpoonup}{p}}^4}
                          {{\stackrel{\rightharpoonup}{p}}^2+4m^2},
$$
\begin{equation}
  2m^2-\omega_lp^0-|\stackrel{\rightharpoonup}{l}||\stackrel{\rightharpoonup}{p}|
  \simeq -{\stackrel{\rightharpoonup}{p}}^2
\end{equation}
\noindent which further lead to
\begin{eqnarray}
B&=&2\lim \limits_{\varepsilon \to 0}
    \delta_{|\stackrel{\rightharpoonup}{l}|,
            |\stackrel{\rightharpoonup}{p}|/2}
   \left\{\int_0^{\infty}dt \ \delta\left[
      \frac{\varepsilon^2{\stackrel{\rightharpoonup}{p}}^4}
                          {{\stackrel{\rightharpoonup}{p}}^2+4m^2}t^2-
           {\stackrel{\rightharpoonup}{p}}^2\right]+
          \int_0^{\infty}dt \ \delta\left[
      \frac{\varepsilon^2{\stackrel{\rightharpoonup}{p}}^4}
                          {{\stackrel{\rightharpoonup}{p}}^2+4m^2}-
           {\stackrel{\rightharpoonup}{p}}^2t^2\right] \right\} \nonumber \\
&=&2\frac{({\stackrel{\rightharpoonup}{p}}^2+4m^2)^{1/2}}
         {{\stackrel{\rightharpoonup}{p}}^2}
         \lim \limits_{\varepsilon \to 0}
         \frac{\delta_{|\stackrel{\rightharpoonup}{l}|,
            |\stackrel{\rightharpoonup}{p}|/2}}
              {\varepsilon |\stackrel{\rightharpoonup}{p}|}
 = 2\frac{({\stackrel{\rightharpoonup}{p}}^2+4m^2)^{1/2}}
         {{\stackrel{\rightharpoonup}{p}}^2}
         \lim \limits_{|\stackrel{\rightharpoonup}{l}| \to
            |\stackrel{\rightharpoonup}{p}|/2 }
         \frac{\delta_{|\stackrel{\rightharpoonup}{l}|,
            |\stackrel{\rightharpoonup}{p}|/2}}
       {\left||\stackrel{\rightharpoonup}{l}| -
       |\stackrel{\rightharpoonup}{p}|/2\right|}   \nonumber \\
 &=&2\frac{({\stackrel{\rightharpoonup}{p}}^2+4m^2)^{1/2}}
         {{\stackrel{\rightharpoonup}{p}}^2}
   \delta(|\stackrel{\rightharpoonup}{l}| -|\stackrel{\rightharpoonup}{p}|/2).
 \end{eqnarray}
 \noindent Substituting Eq. (37) into Eq.(33) we will have
 \begin{equation}
 A(p)|_{p^2=4m^2}=\frac{1}{2|\stackrel{\rightharpoonup}{p}|}
    \left. f(-\sqrt{{\stackrel{\rightharpoonup}{l}}^2+m^2},p^0, \mu)
    \right |_{|\stackrel{\rightharpoonup}{l}|=|\stackrel{\rightharpoonup}{p}|/2}.
\end{equation}
\noindent This means that both $S(p)|_{p^2=4m^2}$ and $R(p)|_{p^2=4m^2}$
with the form of Eq. (38) are finite.  \\
\indent To summarize the above results, we can obtain
\begin{equation}
\lim \limits_{p^2\to 4m^2}(p^2-4m^2)\Gamma_S(p)=0,
\end{equation}
hence $p^2=4m^2$ is not a simple pole of $\Gamma_S(p)$ and $\Gamma_S(p)$
does not represent the propagator for any scalar bound state.  This proves
that no scalar bound state could exist in this model.  This result, seen from
Eqs. (28) and (30), is true at $T=0$, and also maintained at a finite
temperature.  We point out that the above result at $T=0$ is different from the
one obtained in Reference [5].\\
\indent In conclusion, in 3 dimension Gross-Neveu model,
the fermion codensates induced by the scalar four-fermion interactions will
spontaneously break the discrete $Z_2$, the special parity ${\cal P}_1$,
${\cal P}_2$ and the time-revesal symmetries and lead to generation of
dynamical fermion mass. By means of the real-time thermal field theory in the
fermionic bubble diagram approximation, it is clearly shown that the dynamical
fermion mass $m$, as the order parameter
of symmetry breakings, will decrease as temperature increases and finally
becomes zero at some critical temperature $T_c$. This implies restorations of
these discrete symmetries at $T>T_c$. The equation of the criticality curve
of chemical potential and temperature will be indepedent of the momemtum
cut-off and identical to one obtained by the auxialiary scalar field approach
in the $1/N$ expansion if the ratio between the momentum cut-off and the
dynamical fermion mass at $T=0$ is sufficiently large.
Similar to the case in 4D NJL-model, the dynamical mass $m$ has the
$(T_c-T)^{1/2}$ behavior at $T\stackrel{<}{\sim} T_c$ and this fact
indicates the second-order phase transition feature of the symmetry
restorations.  We have also proven by explicit analysis of the fermionic scalar
four-point functions that there is no scalar bound state in this model.\\
\indent The author is grateful to Prof. Zhao Bao-Heng for helpful discussions.
This work was done (in part) in the frame of Associate Membership Programme of
the Abdus Salam International Centre for Theoretical Physics, Trieste, Italy
and partially supported by National Natural Science Foundation of China and by
Grant No. LWTZ-1298 of Chinese Academy of Sciences.

\end{document}